\documentclass[aps,prd,twocolumn,showpacs,superscriptaddress,groupedaddress]{revtex4}  

\usepackage{graphicx}
\usepackage{dcolumn}
\usepackage{bm}
\usepackage{amssymb}  
\usepackage{amsmath}
\usepackage{hyperref} 
\usepackage{color}
\usepackage{enumitem}

\newcommand{\be}{\begin{equation}}
\newcommand{\ee}{\end{equation}}
\newcommand{\bea}{\begin{eqnarray}}
\newcommand{\eea}{\end{eqnarray}}
\newcommand{\nn}{\nonumber}
\newcommand{\mbf}[1][k]{\mathbf{#1}}

\begin{document}

\vspace*{7mm}

\title{Loop corrections to primordial non-Gaussianity}

\author{Sibel Boran}
\email{borans@itu.edu.tr}
\affiliation{Department of Physics, Istanbul Technical University, Maslak 34469 Istanbul, Turkey}

\author{E. O. Kahya}
\email{eokahya@itu.edu.tr}
\affiliation{Department of Physics, Istanbul Technical University, Maslak 34469 Istanbul, Turkey}

\date{\today}
\begin{abstract}
We discuss quantum gravitational loop effects to observable quantities such as curvature power spectrum 
and primordial non-Gaussianity of Cosmic Microwave Background (CMB) radiation. 
We first review the previously shown case where one gets a time-dependence for zeta-zeta correlator 
due to loop corrections. Then we investigate the effect of loop corrections to primordial non-Gaussianity of CMB. 
We conclude that, even with a single scalar inflaton, one might get a huge value for non-Gaussianity 
which would exceed the observed value by at least 30 orders of magnitude. 
Finally we discuss the consequences of this result for scalar driven inflationary models.
\end{abstract}

\maketitle

\section{Introduction}\label{introduction}

Most probably, the founders of quantum gravity did not have high hopes that what they were doing would some day
be tested or even have observational consequences. The CMB power spectrum opened that window to us and now
we can name cosmological perturbations as first quantum gravitational observables that were predicted 
by Mukhanov and Chibisov \cite{MC} for the scalar part and by Starobinsky \cite{Sta} for the tensor part.

Since we already measured the lowest order effect in perturbation theory, the next logical step in any quantum
field theory calculation is to go beyond this level, so-called the tree-level. This is mostly done in order to make
precision tests of a particular model. Although one would expect only small corrections to already known physics
(precision predictions) by calculating these higher order terms in perturbation theory, i.e. loops, new phenomena
beyond our expectations can arise as was the case in the famous one-loop beta function calculation in quantum chromodynamics.

For the past fifteen years, there have been many efforts towards understanding loops in cosmology. 
Among this relatively large literature, the most influential works were those of Weinberg \cite{SW1,SW2}. 
In one of these works, Weinberg asserted a theorem \cite{SW2} related to quantum loop effects in
cosmology: in $N$-th order perturbation theory, quantum corrections can at most be of order $(\alpha\ln a(t))^N$, 
where $\alpha$ is the loop counting parameter and $a(t)$ is the scale factor.

There were plenty of discussions \cite{LSMZ1, KOW1, LSMZ2, LSMZ3, LSMZ4} not about the existence 
but the type of ``infrared logarithm" that arise from the quantum contributions in cosmological correlations.
Two types of infrared logarithm factors that appeared are time-dependent and time-independent. 
The obvious enticing aspect of the time-dependent infrared logarithm is that it grows with time. 
Having this case, the smallness of the loop counting parameter in quantum gravity gets 
counterbalanced by time-dependent infrared logarithms. If one assumes that we observe 50 $e$-folds of inflationary
era, this time-dependent enhancement would only bring a factor of 50. Therefore there is not much hope of observing
this effect any time soon.

Most of the discussions were around the quantum corrections to two-point correlation functions, i.e. power
spectrum; since it is a quantity which is measured more accurately. The debate between the time-dependence and
the time-independence camp went on and both parties published explicit calculations to support their claims
\cite{LSMZ1, KOW1, LSMZ2, LSMZ3, LSMZ4}. The author of this work also contributed to this
discussion claiming the time-dependence. In this work, we do not want to discuss the strengths and weaknesses of each
point of view, but rather want to point out that the ``small" time-dependent loop effects might not be that small; 
if one looks at higher orders of correlation functions such as three-point functions and even higher. The loop effects to
the three-point function were discussed in two separate works. The first one is the work of Giddings and Sloth
where they assumed the semiclassical approximation holds \cite{sg}. The second work is the work of Cogollo {\it et al.} \cite{crv}
where there is an extra scalar field which induces the time-dependent effect. In this work, we will calculate the loop
corrections to external legs of the three-point function using the Hartree approximation. We will show that the loop
corrections have the potential to dominate the tree-level term by 30 orders of magnitude and therefore the perturbation
theory will break down. If this effect survives after a nonperturbative analysis, this would immediately lead to
ruling out all single field inflationary models due to observational constraints on non-Gaussianity of the primordial 
curvature perturbation \cite{Planck1}. This claim might appear very odd, if one naively looks at the above theorem
of Weinberg. But it turns out that the constancy of the tree-level mode function and time-dependency of loop
corrections determine the faith of their contribution to the three-point function, bispectrum.

This paper is organized as follows. In Sec. \ref{s2} we review the case of the time-dependent zeta-zeta correlator arising
from self-interaction of zeta at one-loop order. In Sec. \ref{s3} we use this one-loop corrected time-dependent mode
function to calculate the one-loop corrected primordial
bispectrum. We give our conclusions in Sec. \ref{s4}.

\section{Slow-roll inflation and time-dependent zeta-zeta correlator}\label{s2}

The action of the model that we would like to consider consists of a scalar field $\phi$, an Einstein-Hilbert part and a standard kinetic term 
\bea\label{s}
S &=& \int d^4x \sqrt{-g } \Bigl[\frac{R}{16 \pi G} - \frac12 \phi_{,\mu} \phi_{,\nu} g^{\mu\nu} - V(\phi)\Bigr]\;.
\eea
The metric is that of the Friedmann-Robertson-Walker one which our universe seems to prefer,
\bea
ds^2 &=& -dt^2 + a^2(t) d\vec{x} \cdot d\vec{x} 
\Rightarrow H \equiv \frac{\dot{a}}{a}\;.
\eea
We choose the background inflaton field to be constant at equal-time hypersurfaces as Maldacena \cite{JM} and Weinberg \cite{SW1},
\bea
\phi(t,\vec{x}) - \phi_0(t) &=& 0\;.\label{G0}
\eea

The other $(D-1)$ conditions come from defining 
the unimodular part of the metric $\widetilde{g}_{ij}$,
\bea
g_{ij} &=& a^2(t) e^{2 \zeta(t,\vec{x})} \tilde{g}_{ij}(t,\vec{x}) 
\Rightarrow \sqrt{g} = a^{D-1} e^{(D-1) \zeta}\;.\label{conf}
\eea
By choosing the gauge in the above manner we switched from inflaton field, 
which is the dynamical variable $\phi$ of our theory, to $\zeta$ now parametrizing scalar fluctuations.
Using Einstein's equations for the background scalar field $\phi$, 
one can express its time-derivative in terms of the Hubble parameter as
\bea
\dot{\phi^2} &=& - \frac{\dot{H}}{4 \pi G}\;.
\eea
Another important quantity, called the first slow-roll parameter $\epsilon$, is defined as
\bea
\epsilon \equiv -\frac{\dot{H}}{H^2}  \ll 1\;.
\eea

The next step is using perturbation theory for small fluctuations of the scalar and tensor fields. 
It became customary to use Arnowitt-Deser-Misner formalism to get the quadratic, 
cubic and even higher order parts of the action. The quadratic part of the action for $\zeta$ is \cite{JM}
\bea 
S_{\zeta}^{(2)} &=& \frac{1}{8 \pi G} \int d^4 x \epsilon [a^3 \dot \zeta^2 - a (\partial \zeta)^2]\;
\eea
and the cubic part of zeta is
\bea\label{cubic} 
S_{\zeta}^{(3)} &=& \frac{1}{2 \pi G} \int d^4 x  \epsilon^2 a^5 H \dot \zeta^2 \partial^{-2} \dot \zeta \equiv \int d t L_3(t)\;.
\eea

One can vary the quadratic part of the action and equate to zero, to get the equation of motion for $\zeta$ as
\bea\label{eom}
- \partial_t ( a^{ 3}\; \epsilon \; \dot \zeta ) + a  \epsilon  \partial^2 \zeta &=& 0\;.
\eea
The standard way of solving this equation for quantum fields is going into momentum space and expressing $\zeta$ as a mode sum 
\bea\label{mod}
\zeta_{\mbf[k]}(t) &=& u_k(t) a_{\mbf} + u^*_k(t) a^\dagger_{-\mbf}\;,
\eea
where $a^\dagger_{-\mbf}$ and $a_{\mbf}$ are creation and annihilation operators that obey canonical quantization conditions. 

It is best to go to conformal time to write the expression for the mode function,
\bea
d \eta \equiv -a dt \Rightarrow ds^2 &=&-dt^2 + a^2(t)  d\vec{x} \cdot d\vec{x}\nn\\
&=& a^2(t)(-d\eta^2 + d\vec{x} \cdot d\vec{x})\;,
\eea 
so that the geometry is conformally flat. Another motivation for choosing conformal time 
is related to the fact that there is not a unique choice of vacuum in curved space. 
One takes the expression \eqref{mod} and uses that to solve equation \eqref{eom} for $u_k(\eta)$, 
\bea
u_k(\eta) &=&  \frac{H}{\sqrt{2 \epsilon}} \frac{1}{\sqrt{2 k^3}} (1+ik \eta) e^{-ik \eta}\;,
\eea
which corresponds to the positive frequency modes. 
By choosing conformal time for coordinate system, it is easy to see that this solution 
for the mode function behaves like Minkowski in the early time limit. 
This solution is called Bunch-Davies vacuum solution \cite{BD}.

Let us define the curvature power spectrum:
\bea
\Delta^2_{\mathcal{R}}(k,t) \equiv \frac{k^3}{2 \pi^2} \int \!\! d^3x \,
e^{-i \vec{k} \cdot \vec{x}} \Bigl\langle \Omega \Bigl\vert \mathcal{R}(t,\vec{0} )
\mathcal{R}(t,\vec{x}) \Bigr\vert \Omega \Bigr\rangle \;, 
\eea
where $\mathcal{R}$ is related to the 3-curvature and is equal to $\zeta$ at the linearized order \cite{LL};
\bea\label{Rdef}
\mathcal{R}(t,\vec{x}) \equiv -\frac{a^2(t)}{4 \nabla^2} R 
&=& \zeta(t,\vec{x}) + O\Bigl(\zeta^2,\zeta h, h^2\Bigr) \;. 
\eea
Therefore curvature power spectrum also goes by the name ``zeta-zeta correlator'' as well. 
The latest value of the curvature power spectrum constructed from measurements is \cite{Planck2}
\bea
\Delta^2_{\mathcal{R}}(k)=\Bigl(2.198^{+0.076}_{-0.085}\Bigr) 
\times 10^{-9} \Bigl( \frac{k}{0.002~{\rm Mpc}^{-1}}\Bigr)^{-0.0345 \pm 0.0062}.\label{data}\nn\\
\eea
The theoretical prediction at tree-level gives us
\bea
\Bigl[\Delta^2_{\mathcal{R}}(k) \Bigr]_{\rm tree} 
\approx \frac{4 G k^3}{\pi} \times \vert u(t,k)\vert^2 
\approx \frac{G H^2(t_k)}{\pi \epsilon}\;.
\eea
Although one would expect that tree-order quantum gravity calculations capture the full effect, 
it is natural to wonder what happens beyond that. Therefore we want to know if loop corrections to this measurable quantity make 
any difference. There have been many efforts to answer this questions in the past ten years 
\cite{TY1, KKT, TY2, GS1, GS2, KOW2, AEL, See, BVS, wy, nea, en}.

\begin{figure}[htbp]
\centering
\includegraphics[height=20mm]{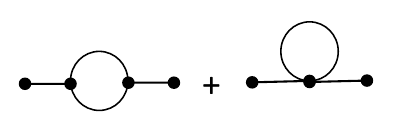}
\caption{One-loop correction is sourced by cubic and quartic self-interactions of $\zeta$.}
\label{twoloops}
\end{figure}

In a particular curious case \cite{KOW1} it was shown that one can get an enhanced time-dependent 
correction to $\zeta$-$\zeta$ correlator at one-loop order coming from the Feynman diagrams in Fig. \ref{twoloops}. 

The one-loop corrected curvature power spectrum gives
\bea\label{selfcor}
\Bigl[\Delta^2_{\mathcal{R}}(k,t) \Bigr]_{\zeta{\rm loops}} 
\approx \frac{G H^2}{\pi \epsilon} \Bigl\{ 1+ \frac{27 G H^2}{4 \pi \epsilon}
\ln(a) + O(G^2 H^4) \Bigr\}\;.\nn\\ 
\eea
This corresponds to a correction to the tree-level scalar mode function as
\bea
\Rightarrow u(t,k) \approx \frac{H}{\sqrt{2\epsilon}} \frac{1}{\sqrt{2 k^3}} 
\Bigl\{1 + \frac{27 G H^2}{8 \pi \epsilon} \, \ln(a) \Bigr\} \; , \qquad
\eea
if one uses Hartree approximation.

The $\zeta$-$\zeta$ correlator becomes time-dependent if there is least one undifferentiated field 
in the action at the relevant order. These so-called infrared logarithms, as well as $1/\epsilon$ term, 
enhance this one-loop effect by 3 orders of magnitude. But the smallness of $GH^2\approx 10^{-10}$ 
overshadows this enhancement and makes the total one-loop correction to be at most at the order of $10^{-6}$. 
The degree of the precision of the current experiments is well below the necessary level to untangle 
this one-loop effect. But still the effect is not hopelessly small. 

For the last five years there has been some discussion about the time-dependence of the $\zeta$-$\zeta$ correlator. 
It has even been claimed \cite{LSMZ4} that this quantity is constant at all loops, 
which we find to be highly dubious since even at tree-level it only asymptotes to a constant. 
The point of this work is not to argue the time-dependence of $\zeta$-$\zeta$ correlator, 
but rather go towards another direction; which is loop corrections to three-point function 
and to see what the consequences of time dependence of $\zeta$ are. 
We believe that the real enhancement of time-dependent $\zeta$-$\zeta$ correlator arises 
if one calculates three-point function for $\zeta$. 
It turns out the one-loop correction to this quantity might totally dominate the tree-level result. 
Therefore when searching for enhanced quantum gravity corrections, 
the more interesting quantity to calculate is the three-point function, which is the subject of the next section. 

\section{Bispectrum at tree-level}\label{s3}

One can write the primordial bispectrum in terms of the Fourier transformed three-point function as 
\bea
\langle\zeta_{\bold{k_1}}\zeta_{\bold{k_2}}\zeta_{\bold{k_3}}\rangle
=(2\pi)^3 \delta^3(\bold{k_1}+\bold{k_2}+\bold{k_3}) B_{\zeta}(k_1,k_2,k_3)\;.
\eea  
Assuming a local form for the bispectrum where the non-Gaussian $\zeta$ field 
is produced from the Gaussian background $\zeta_{g}$ field as
\bea
\zeta(\mbf[x]) &=& \zeta_g(\mbf[x]) + (3/5) f_{\text{NL}} \zeta_g^2(\mbf[x]) + \mathcal{O}(\zeta_g^3)\;.
\eea

One can show that the bispectrum peaks at the so called ``squeezed" triangle, 
for which one takes one wave number much smaller than the other two, i.e. $k_1 \approx k_2 \gg k_3$. 
For the case of squeezed limit bispectrum can be expressed in terms of power spectrum with the following equation
\bea\label{bispec3}
B_\zeta^{\rm{local}}(k_1,k_2,k_3)_{k_1 \approx k_2 \gg k_3}\approx\frac{12}{5}f_{\text{NL}}P(k_1)P(k_3)\;,
\eea
where the late time limit of the power spectrum is
\bea
P(k)&=& |u_k|^{2}_{\eta \rightarrow 0} \;=\; \frac{H^2}{2 \epsilon} \; \frac1{2  k^3}\;.
\eea
If Creminelli-Zaldarriaga consistency \cite{cz} condition for single field inflation models hold, 
the bispectrum in the local limit (or squeezed-limit) can be written as
\bea\label{bispce1}
B_{\zeta} (k_3 \ll k_1)\;=\; (1-n_s) P(k_1) P(k_3)\;,
\eea
where $n_s(k)$ is called the spectral tilt index and is defined as
\bea\label{sti}
n_s -1 \equiv \frac{d ln (P(k))}{d ln (k)}\;.
\eea

Instead of expressing the three-point function in terms of two-point functions, 
one can directly compute the in-in expectation value of $\zeta$ and use \eqref{cubic} 
for the interaction Hamiltonian and can get the following expression for the bispectrum \cite{ganc}:
\bea\label{bispce2}
B_{\zeta} (k_1,k_2,k_3) &=& 8i \frac{\epsilon^2}{H^2} \sum_{k_i} (\frac{1}{k_i^2}) u_{k_1}(\bar{\eta})u_{k_2}(\bar{\eta})u_{k_3}(\bar{\eta}) \nn\\
& &\times \int_{\eta_0}^{\bar{\eta}} d\eta \frac1{\eta^3} u_{k_1}^{'*}u_{k_2}^{'*}u_{k_3}^{'*} +c.c.\;.
\eea   
The main point of this calculation is the integral that we have in the above expression for the bispectrum,
\bea\label{bisint}
\int_{\eta_{0}}^{\bar{\eta}} d\eta \frac{1}{\eta^{3}} \tilde{u}_{k_1}^{'*}\tilde{u}_{k_2}^{'*}\tilde{u}_{k_3}^{'*}\;.
\eea
If we take the tree-order mode function for the above expression it is obvious that 
we will get a small non-Gaussianity. This is due to the fact that the three-point
function, therefore non-Gaussianity, is proportional to the change of the mode function 
for each wave number $k_1, k_2, k_3$. Since for each mode the mode function itself goes to a
constant after the horizon crossing, the change of those tree-level mode functions will be very small. 
On the other hand, the two-point function, therefore power spectrum, is proportional to 
the magnitude of the mode function. Let us highlight this point by giving equations:
\bea\label{twothree}
{\rm power\;spectrum} &\sim& \langle\zeta_{\bold{k_1}}\zeta_{\bold{k_2}}\rangle \sim  \delta^3(\bold{k_1}+\bold{k_2})|u_k|^{2}\nn\\
{\rm non-Gaussianity} &\sim& \langle\zeta_{\bold{k_1}}\zeta_{\bold{k_2}}\zeta_{\bold{k_3}}\rangle\nn\\ 
&\sim & \delta^3(\bold{k_1}+\bold{k_2}+\bold{k_3})  {u}_{k_1}(\bar{\eta}){u}_{k_2}(\bar{\eta}){u}_{k_3}(\bar{\eta})\nn\\
&\times & \int_{\eta_0}^{\bar{\eta}} d\eta \frac1{\eta^3} {u}_{k_1}^{'*}{u}_{k_2}^{'*}{u}_{k_3}^{'*} +c.c.
\eea

\section{One-loop correction to bispectrum: a particular example}\label{s4}

We would like to find the answer to the following question: How big is the effect of loops to n-point
functions of primordial curvature perturbation during Inflation? The answer of the above question for two-point
and three-point functions is are related to the magnitude and the time derivative of the scalar mode functions,
respectively.

One should do the full computation to give a definitive answer of time dependence of bispectrum. 
In this work, our aim is to show that it is possible to get huge enhancements
at one-loop order and for that we would like to consider the simplest case where the external legs are corrected

\begin{figure}[htbp]
\centering
\includegraphics[height=25mm]{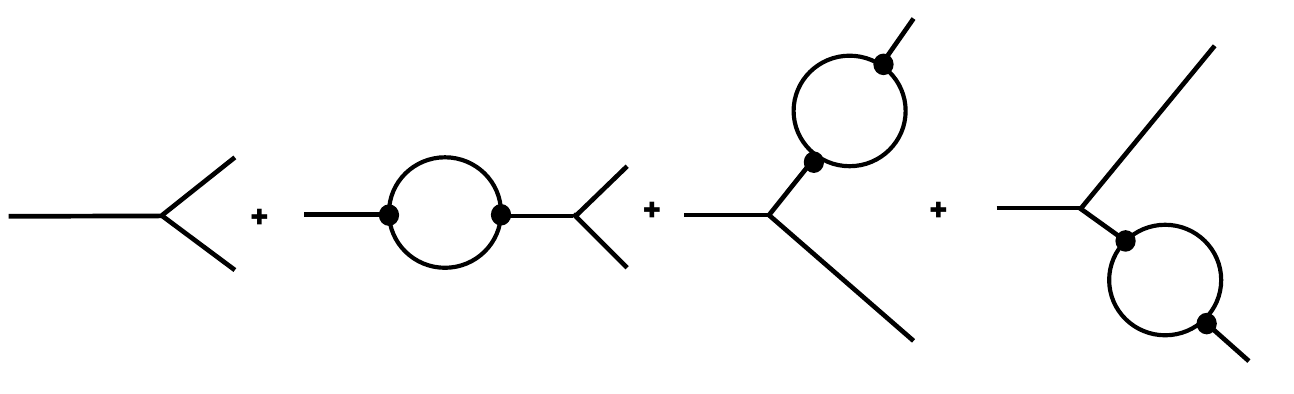}
\caption{Bispectrum: Tree-level and corrections to external legs.}
\label{f3}
\end{figure}

Here we would like to make a further simplification, namely use the form of the one-loop corrected mode
function \eqref{twothree} where we applied a Hartree approximation.

The one-loop corrected mode function expression and its derivative with respect to conformal time and their long
wavelength limits ($k \eta \ll 1 $) are
\bea
u_{tree}&=&\frac{H}{\sqrt{2 \epsilon}} \frac{1}{\sqrt{2 k^3}}(1+ik \eta) e^{-ik \eta} \nn\\ 
&\Rightarrow & \frac{H}{\sqrt{2 }} \frac{1}{\sqrt{2 k^3}}  \Biggl\{1\;+  \frac{k^2 \eta^2}{2}+ ... \Biggr\}\;,
\eea
\bea
{u}_{\rm 1-loop} &\Rightarrow&  \frac{H}{\sqrt{2\epsilon}} \frac{1}{\sqrt{2 k^3}} \Bigl\{1 + O(1) G H^2 \ln(a) \Bigr\}\nn\\
&=& u_{tree} \Bigl\{1 + O(1)\,\, G H^2 \, \ln(a) \Bigr\}\;,
\eea
\bea
u_{tree}^{'}  &\Rightarrow &  \frac{H}{\sqrt{2 \epsilon}} \frac{1}{\sqrt{2 k^3}}  \Biggl\{k^2 \eta+ ... \Biggr\}\;,
\eea
\bea
{u}_{\rm 1-loop}^{'} &\Rightarrow & \frac{H}{\sqrt{2\epsilon}} \frac{1}{\sqrt{2 k^3}}  k^2 \eta \Bigl\{1 + O(1) G H^2 \frac{1}{k^2 \eta^2} \Bigr\}\nn\\
&=& u_{tree}^{'} \Bigl\{1 + O(1)  G H^2  \frac{1}{k^2 \eta^2} + ... \Bigr\}\;.
\eea
The difference between the one-loop corrected mode function's and tree-level mode function's 
time derivative is of the order of $G H^2$ $\approx10^{-10}$ as expected, 
but also multiplied with an additional factor of $1/k^2 \eta^2$. 
For the super-horizon modes ($k \eta \ll 1 $) with the relevant 50 $e$-folds 
this brings an extra factor of $10^{20}$ which makes the one-loop correction 
to dwarf the tree-level part of the mode function. 

The mathematical reasons for this huge effect is the following. 
The derivative with respect to conformal time brings an extra factor of $H a(t)$ when it acts on a power of  $a(t)$. 
Since the tree-level mode function is constant after horizon crossing, 
this time derivative does no good to boost the leading term, it simply annuls it. 
But for the case one-loop corrected mode function, the time derivative acts on the $\ln a$ 
and gives a chance to the constant leading term of the tree-level part of the mode function to survive. 
Not only does it survive, but also it gets boosted by the extra factor of $H a(t)$. 

Since the integral that appear in the the non-Gaussianity ($\zeta \zeta \zeta$ correlator) \eqref{twothree} 
has three factors of ${u}_{\rm 1-loop}^{'}$ the one-loop correction to the three-point function is 
30 orders magnitude bigger than the tree-level term. 
Since the one-loop correction to $P(k)$ terms are very small, 
this extremely large value of bispectrum can only be achieved by having a huge $f_{\rm NL}$ parameter, 
if we look at equation \eqref{bispec3}. It also implies that Creminelli-Zaldarriaga 
consistency condition for single scalar field is not satisfied here, since equation \eqref{bispce1} 
could not be satisfied with a bispectrum this big. 
However this does not mean the invalidation of the consistency condition, since the condition assumes 
time-independency a priori, although there are cases where the condition is violated \cite{rms, nfs}. 
But most importantly an $f_{\rm NL}$ parameter of this magnitude results into ruling out 
all single field inflation models due to the observational limits on the non-Gaussianity parameter. 

We would like to point out that we are not claiming Weinberg's theorem is incorrect. 
Mathematically what happens is that, the time derivative acting on the constant term, 
which is the leading term in the long wavelength expansion of scalar mode function kills it. 
On the other hand, the quantum corrected time-dependent mode functions long wavelength expansion 
leave a room for the constant term to survive by letting the time derivative hit on the infrared logarithm.  
Due to this wondrous interaction, these one-loop terms dominate the tree-level term by 30 orders of magnitude, 
which results in breaking down of the perturbation theory. Therefore as Weinberg pointed out \cite{SW2} 
we see the need of a non-perturbative method for cosmological correlations.

The integrals that appear in \eqref{twothree} can be evaluated \cite{BK} analytically in a general vacuum choice, 
called non-Bunch-Davies initial state which would lead to an increase of the non-Gaussianity parameter $f_{\rm NL}$ 
even at the tree-level \cite{ganc,ap}. For the case of non-Bunch-Davies initial state the one-loop 
term again dominates the tree-level one and be ruled out as well.

\section{Discussion}\label{s5}

The success of inflationary cosmology is appalling. 
This simple idea solves homogeneity, flatness, horizon, isotropy and primordial monopole problems of 
standard cosmology with a single shot \cite{guth,linde,bst}. 
With inflation, linking quantum physics with cosmology, we can understand 
the origin of all matter from primordial quantum fluctuations. 
Getting first quantum gravitational data, such as curvature power spectrum 
with small error bars is a major success in itself. 

It is therefore time to go beyond this tree-level effect and investigate possible consequences, 
which we can name as precision inflationary cosmology. Towards this direction, 
one possible thing to do is calculating loop corrections to cosmological correlations. 
At first look, one would naturally think that this is a futile effort due to the smallness 
of the loop counting parameter $GH^2$. But still there was a lot of attention to loop corrections 
to power spectrum despite the smallness of them.

Cosmological loop corrections bring a typical infrared logarithm and are divided into 
three categories according to the form of the logarithmic factors: $Log(H\mu)$, $Log(kL)$ and  $Log(a(t))$ \cite{LSMZ3}. 
The first case is claimed to be due to making an error in the implementing diffeomorphism invariant regularization 
and the second being a projection effect that will be removed if one computes observable quantities. 
The final case is also dismissed in the mentioned work on the grounds of symmetry arguments 
as well as extrapolating this time-dependent effect to reheating and baryogenesis 
and claiming that predictivity of inflation will be lost. 
One can certainly reply to the above criticisms and perhaps one should. 
But in this work, we would like to bring a different viewpoint to this discussion that is more dramatic. 

First of all, time-dependent zeta do occur even without loop corrections, 
such as multifield inflationary models and entropy perturbations. 
The time-dependence that we are interested in, that has the form of $Log(a(t))$, 
are originated from loop corrections. We investigated the minimal case where the only scalar field is the inflaton.
We first reviewed time-dependent loop corrections to two-point functions, i.e. power spectrum,  
which arises due to $\zeta$-$\zeta$ self interactions at one-loop order. 
In principle they are important; on the other hand, from an observational perspective they are irrelevant at the moment.
We took the one-loop corrected scalar mode function and used that to calculate 
the loop corrections to external legs of the three-point function using Hartree approximation. 
We concluded that they grow with the square of the scale factor. We showed that the loop corrections 
have the potential of dominating the tree-level term by 30 orders of magnitude and lead 
to breaking down of the perturbation theory. If this effect survives after a nonperturbative analysis, 
that would result into an immediate ruling out all single-scalar driven models of inflation.

Therefore, non-Gaussianity is a better place, compared to power spectrum, to look for quantum gravitational corrections. 
The technical reason of this is: 
\begin{enumerate}
\item{Power spectrum is related to the magnitude of mode function. 
\\ Therefore it goes like : constant (tree-level) +  a small correction (loops)}
\item{Non-Gaussianity is related to the time derivative of the mode function. 
\\ Therefore it goes like : almost zero (tree-level) +  a not so small correction(loops) compared to zero} 
\end{enumerate}  
Therefore the real treasure is hidden in the higher order correlation functions, not in the power spectrum. 
We also showed that this would imply a huge ($10^{20}$ times bigger than tree-level prediction) 
non-Gaussianity $f_{\rm NL}$ parameter, leading to an immediate contradiction 
with the constraints on observed value of $f_{\rm NL}$ parameter. 
 
\begin{figure}[htbp]
\centering
\includegraphics[height=25mm]{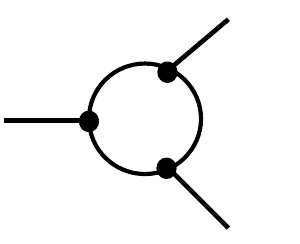}
\caption{The remaining diagram at one-loop order.}
\label{f4}
\end{figure}

At this point we would like to discuss the possible ways of avoiding huge $f_{\rm NL}$ parameter. 
Let us remember that there is another type of diagram that will contribute at one-loop order to three-point function. 
It might turn out that this vertex correction (Fig. \ref{f4}) cancel the total contribution coming from Fig. \ref{f3}.

For that end it is useful to look at the work of Cogolo {\it et al.} \cite{crv}, 
where they show that this kind of diagram dominates the whole series of diagrams. 
They consider and extra scalar field and the effect arises due to that; 
but still a similar thing might happen for the single scalar field situation.

We want to conclude the discussion section by giving six points that should further be investigated in detail which  might change the picture:
\begin{enumerate}[label=(\roman*)]
  \item Vertex correction
  \item Hartree approximation
  \item Single scalar field (inflaton), adiabatic perturbations
  \item Time-dependent $\zeta$-$\zeta$ correlator from loops
  \item Almost constant slow-roll parameter $\epsilon$
  \item A nonperturbative analysis.
\end{enumerate}
The first possibility is the vertex correction exactly cancelling the effects given above. 
It might be that using Hartree approximation is the source of the mentioned effect. 
For that, one should do the full calculation and see if the effect is not there.
One can imagine scenarios where a spectator field causing a similar effect which might cancel the $\zeta$ loops. 
This only happens for particular situations \cite{crv}. 
One could also try to incorporate the time-dependence of the $\epsilon$ parameter and investigate the consequences of that.

Our work highlights the need of a nonperturbative method to cosmological correlations. 
A nonperturbative method was found by Starobinsky and Yokoyoma \cite{sy} for self-interacting scalar fields 
and Tsamis and Woodard for Scalar Quantum Electrodynamics \cite{tw}, where they were able to resum the leading infrared log terms 
in the whole perturbative series. And it might turn out that, one can avoid a big $f_{\rm NL}$ parameter 
after calculating the quantum effects using a nonperturbative method. 
One can simply say that it is the fourth assumption that is wrong and maybe it is so. 
But no matter what the solution is, quantum loop corrections that result into 
time-dependent scalar mode functions have consequences that are so important 
and will be of such magnitude that they can not be swept under the rug. 

We would like to end our discussion by pointing out to a curious work done by Pattison {\it et al.} \cite{PVAW}, 
where the probability density function (PDF) of curvature perturbations were calculated by using stochastic $\delta N$ formalism. 
During this period, due to quantum diffusion effects, stochastic force would determine the inflaton dynamics 
and PDF of $\zeta$ has the form elliptic theta functions. 
They claim that, in the limit where the potential is exactly flat 
and stochastic effects dominate, one gets highly non-Gaussian curvature perturbations. 
This claim, which might be an artefact of using stochastic $\delta N$ formalism, 
should certainly be checked by making rigorous loop calculations. 
This work also shows the need for doing a one-loop calculation 
(fully renormalized with all the relevant interactions included) 
of the three-point function in particular and n-point functions of $\zeta$ in general.



\end{document}